# Trust-Aware Control of Automated Vehicles in Car-Following Interactions with Human Drivers

Mehmet Fatih Ozkan and Yao Ma

*Abstract*— Trust is essential for automated vehicles (AVs) to promote and sustain technology acceptance in human-dominated traffic scenarios. However, computational trust dynamic models describing the interactive relationship between the AVs and surrounding human drivers in traffic rarely exist. This paper aims to fill this gap by developing a quantitative trust dynamic model of the human driver in the car-following interaction with the AV and incorporating the proposed trust dynamic model into the AV's control design. The human driver's trust level is modeled as a plan evaluation metric that measures the explicability of the AV's plan from the human driver's perspective, and the explicability score of the AV's plan is integrated into the AV's decision-making process. With the proposed approach, trust-aware AVs generate explicable plans by optimizing both predefined plans and explicability of the plans in the car-following interactions with the following human driver. The results collectively demonstrate that the trust-aware AV can generate more explicable plans and achieve a higher trust level for the human driver compared to trust-unaware AV in human-AV interactions.

## I. INTRODUCTION

It has been shown that automated vehicle (AV) technologies can improve the efficiency and sustainability of human-dominated mixed traffic through proper interactions with surrounding human drivers [1]. However, such benefits may not be achieved if human drivers lack trust in AV technologies. Considering car-following interactions between human-driven vehicles and AVs, the current AV longitudinal motion planning strategies [2-3] do not explicitly account for the influences on the following human driver's trust level. This may cause AVs to generate undesirable behaviors that are perceived as out of expectations of the human drivers, and such AVs' unexpected behaviors can deteriorate the trust level of the human drivers. Therefore, the development of AV systems should aim to meet the expectations of the human drivers and improve the human driver's trust level, resulting in beneficial human-AV interactions. To accomplish such an objective, we propose developing a trust-aware AV control approach that incorporates the trust level of the human driver in the car-following interaction with the AV. This is achieved by modeling the trust level of the human driver as a plan evaluation index that quantifies the explicability of the AV's plan from a human driver's perspective rather than the inherent value of the human's cognitive trust level that forms in the human's mind which is difficult to quantify.

M. F. Ozkan and Y. Ma are with the Department of Mechanical Engineering, Texas Tech University (e-mail: mehmet.ozkan@ttu.edu and yao.ma@ttu.edu).

In human-automation interactive systems, trust has been defined as the attitude towards automation where automation will help achieve a human's goal in a situation characterized by uncertainty and vulnerability [4]. Human trust has a dynamic nature as it can change during the interaction with the other agents [5], and the difference between the actual performance of the automation and the human's expectations can significantly affect the trust level, as stated in [6]. Hence, a quantitative dynamic model is necessary to understand the human's dynamic trust level and fulfill the human's expectations in human-automation interactive systems.

Existing trust models for human-AV interactive systems are mostly studied at the level of driver-to-vehicle applications, and most of these trust models are qualitative without considering the quantitative dynamic model of human trust [7-8]. A quantitative dynamic trust model for the driver-to-vehicle application in human-AV interaction was developed for the first time in [6]. The proposed computational trust dynamic model describes the driver's trust in the adaptive cruise control (ACC) system and can be integrated into the trust-aware control design of the AV to meet the human's expectations and increase the trust level of the driver. On the other hand, no quantitative trust dynamic models and trust-aware control designs for human-AV interactive systems have been developed at the level of vehicle-to-vehicle interactions. To this end, it is crucial to establish a suitable trust dynamic model that defines the trust level of the human driver and a trust-aware control strategy that incorporates the human driver's expectations to generate such trust-aware behaviors in human-AV interactions.

This study aims to design a quantitative dynamic trust model for the following human driver in the car-following interaction with the AV and to integrate the proposed trust model in the trust-aware control design of the AV. The distinctive contributions of this study include the following two aspects: 1) a quantitative dynamic model for the human driver's trust in the car-following interactions with the automated vehicle is proposed. 2) a plan explicability is integrated into the longitudinal motion planning of the automated vehicle to enable trust-aware behaviors in the car-following interactions with the human driver. To the best of the authors' knowledge, this is the first study explicitly addressing the quantitative dynamic trust model and trust-aware control strategy in the human-automated vehicle interactive systems at the level of vehicle-to-vehicle interactions.

The rest of this paper is structured as follows. The problem formulation in human-AV interaction is presented in Section II. Trust-aware behavior planning is introduced in

Section III. The trust-aware control design is developed in Section IV. The performance of the trust-aware control design in realistic driving scenarios is investigated using numerical simulations in Section V. At last, final remarks are made in Section VI.

## II. PROBLEM FORMULATION

In this work, we explore an interactive system between the AV and the following human driver in the car-following scenario, as shown in Fig. 1. The goal is to develop a trust-aware AV that considers both the AV's original longitudinal motion plan and the explicability of the AV's plan in the human driver's mind.

We consider that AV $(\mathcal{R})$ and the human driver $(\mathcal{H})$ have unique models $\mathcal{M}_\mathcal{R}$ and $\mathcal{M}_\mathcal{H}$ to construct their plans in the longitudinal driving scenario, respectively. We further assume that AV and the human driver are rational agents who aim to make decisions in the interests of their cost functions across their planning horizons. Accordingly, it can be considered that the longitudinal motion control of the AV and human driver can be addressed by employing the Model Predictive Control (MPC) strategy over the planning horizon. Let $C_{\mathcal{M}_i}$ be the cost function of the AV and human driver $(i \in \{\mathcal{R}, \mathcal{H}\})$ across the planning horizon $N$:

$$C_{\mathcal{M}_i}\left(x^t_{\mathcal{M}_i}, \mathbf{u}_{\mathcal{M}_i}, \mathbf{u}_{\mathcal{M}_\mathcal{R}^\mathcal{H}}\right) = \sum_{k=0}^{N-1} c_i\left(x^{t,k}_{\mathcal{M}_i}, u^k_{\mathcal{M}_i}, u^k_{\mathcal{M}_\mathcal{R}^\mathcal{H}}\right) \quad (1)$$

where $x^t_{\mathcal{M}_i}$, $x^{t,k}_{\mathcal{M}_i}$ and $\mathbf{u}_{\mathcal{M}_i} = \left(u^0_{\mathcal{M}_i}, u^1_{\mathcal{M}_i}, ..., u^{N-1}_{\mathcal{M}_i}\right)^T$ represent the system state at the time $t$, $(t+k)^{th}$ predicted system state, and the longitudinal motion plan that consists of the sequence of the longitudinal accelerations of the AV and human driver, respectively, and $\mathbf{u}_{\mathcal{M}_\mathcal{R}^\mathcal{H}} = \left(u^0_{\mathcal{M}_\mathcal{R}^\mathcal{H}}, u^1_{\mathcal{M}_\mathcal{R}^\mathcal{H}}, ..., u^{N-1}_{\mathcal{M}_\mathcal{R}^\mathcal{H}}\right)^T$ is the expected AV's plan that includes the sequence of the predicted AV's longitudinal accelerations from the human's perspective.

At every time step $t$, the AV can generate its optimal plan by minimizing $C_{\mathcal{M}_\mathcal{R}}$ and compute its optimal longitudinal acceleration vector:

$$\mathbf{u}^*_{\mathcal{M}_\mathcal{R}} = \arg\min_{\mathbf{u}_{\mathcal{M}_\mathcal{R}}} C_{\mathcal{M}_\mathcal{R}}\left(x^t_{\mathcal{M}_\mathcal{R}}, \mathbf{u}_{\mathcal{M}_\mathcal{R}}, \mathbf{u}_{\mathcal{M}_\mathcal{R}^\mathcal{H}}\right) \quad (2)$$

The human driver can then construct an optimal plan with the interpretation of the AV's plan $\mathbf{u}_{\mathcal{M}_\mathcal{R}^\mathcal{H}}$ by minimizing $C_{\mathcal{M}_\mathcal{H}}$ and compute its optimal longitudinal acceleration vector:

$$\mathbf{u}^*_{\mathcal{M}_\mathcal{H}} = \arg\min_{\mathbf{u}_{\mathcal{M}_\mathcal{H}}} C_{\mathcal{M}_\mathcal{H}}\left(x^t_{\mathcal{M}_\mathcal{H}}, \mathbf{u}_{\mathcal{M}_\mathcal{H}}, \mathbf{u}_{\mathcal{M}_\mathcal{R}^\mathcal{H}}\right) \quad (3)$$

By incorporating the previously specified working scheme and assumptions of the human-AV interactive system, we aim to generate trust-aware behaviors of the AV by including the AV's plan explicability from the human driver's cognitive process into the AV's decision-making problem. The implementation details of the AV's trust-aware behavior planning will be presented in the next section.

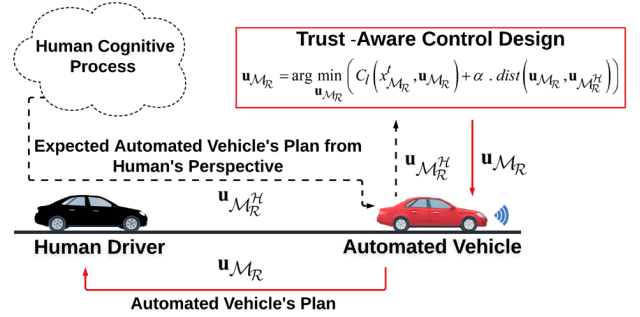

Fig. 1 An illustration of the car-following interaction between the human driver and trust-aware automated vehicle.

## III. TRUST-AWARE BEHAVIOR PLANNING

The AV needs to accomplish a longitudinal motion planning task with respect to its model $\mathcal{M}_\mathcal{R}$ when interacting with the following human driver in the car-following scenario. Meanwhile, the AV also needs to understand its own model $\mathcal{M}_\mathcal{R}$ in the human driver's eyes $\mathcal{M}_\mathcal{R}^\mathcal{H}$ to generate explicable plans in the human-AV interactions. Hence, a plan explicability needs to be determined and integrated as an additional factor into the cost function of the AV. Therefore, the AV's cost function can be expressed as follows, motivated from [9]:

$$C_{\mathcal{M}_\mathcal{R}} = C_l\left(x^t_{\mathcal{M}_\mathcal{R}}, \mathbf{u}_{\mathcal{M}_\mathcal{R}}\right) + \alpha \cdot dist\left(\mathbf{u}_{\mathcal{M}_\mathcal{R}}, \mathbf{u}_{\mathcal{M}_\mathcal{R}^\mathcal{H}}\right) \quad (4)$$

where $C_l$ is the cost function for an AV's original longitudinal motion task; $dist$ defines the distance between the AV's plan $\mathbf{u}_{\mathcal{M}_\mathcal{R}}$ and expected AV's plan from the human drivers' perspective $\mathbf{u}_{\mathcal{M}_\mathcal{R}^\mathcal{H}}$ and $\alpha$ is the relative weight. The goal is to find a plan that minimizes a weighted sum of the cost of the AV's original plan and the differences between the two plans; $\mathbf{u}_{\mathcal{M}_\mathcal{R}}$ and $\mathbf{u}_{\mathcal{M}_\mathcal{R}^\mathcal{H}}$. Given the AV's plan, the challenge arises in computing the $dist$ function in (4). It is worth noting that if $\mathcal{M}_\mathcal{R}^\mathcal{H}$ is known or can be estimated, the only thing left is to choose a suitable $dist$ function. However, $\mathcal{M}_\mathcal{R}^\mathcal{H}$ is generally hidden, difficult to obtain, and can be completely different from $\mathcal{M}_\mathcal{R}$ in practice. In the following sections, we will present the implementation details of the AV's motion planning with respect to $\mathcal{M}_\mathcal{R}$ and provide the methods to estimate $\mathcal{M}_\mathcal{R}^\mathcal{H}$ in the car-following interactions.

### A. Plan Generation

In this study, we formulate the AV's original longitudinal motion plan with the constant time headway policy. The constant time headway policy has been widely adopted as a longitudinal motion planning strategy for AVs, with the aim of maintaining a constant time gap between the AV and its preceding vehicle on the road [10]. Hence, we formulate the

cost function of the AV's original longitudinal motion plan as follows:

$$C_l\left(x^t_{\mathcal{M}_\mathcal{R}}, \mathbf{u}_{\mathcal{M}_\mathcal{R}}\right) = \sum_{k=0}^{N-1}\left(d^k_{CTH} - d^k_{\mathcal{M}_\mathcal{R}}\right)^2, \quad d^k_{CTH} = d_s + v^k_{\mathcal{M}_\mathcal{R}}\tau_{\mathcal{M}_\mathcal{R}}$$
$$\text{s.t. } d^{k+1}_{\mathcal{M}_\mathcal{R}} = d^k_{\mathcal{M}_\mathcal{R}} + v^k_{PT} - v^k_{\mathcal{M}_\mathcal{R}}, \quad v^{k+1}_{\mathcal{M}_\mathcal{R}} = v^k_{\mathcal{M}_\mathcal{R}} + u^k_{\mathcal{M}_\mathcal{R}}$$
(5)

where $d^k_{CTH}$ is the desired car-following distance with constant time headway approach; $d_{\mathcal{M}_\mathcal{R}}$ is the gap distance of the AV to its preceding traffic; $d_s$ is the standstill gap distance; $v_{\mathcal{M}_\mathcal{R}}$ and $v_{PT}$ are the longitudinal speed of the AV and preceding traffic, respectively, and $\tau_{\mathcal{M}_\mathcal{R}}$ is the desired constant time headway.

### B. Plan Estimation

In human-AV interactions, as discussed previously, the AV's model from the human perspective $\mathcal{M}^{\mathcal{H}}_\mathcal{R}$ is naturally concealed, hard to learn, and can be arbitrarily distinctive from the AV's model $\mathcal{M}_\mathcal{R}$. Therefore, our solution is to use such reasonable plan estimation methods to anticipate the AV's plan from the view of the human driver $\mathbf{u}_{\mathcal{M}^{\mathcal{H}}_\mathcal{R}}$ and approximate the value of the *dist* function accordingly. Here, we consider two representative plan estimation methods:

**Time-series motion prediction:** In real-world interactive traffic scenarios, the human drivers usually evaluate the behavior of the other drivers with their past and current observations to estimate the future behavior of those surrounding drivers for building motion plans [11]. Hence, a prediction method that considers the current-past temporal dependency can estimate the AV's plan from the human drivers' perspective in the human-AV interactions. To this end, we use our previously developed gated recurrent unit (GRU) network-based time series prediction model [12] to estimate the AV's longitudinal motion plan over the planning horizon from the human driver's perception. The GRU model is one of the implementations of the recurrent neural network (RNN) structure [13], and it has the advantage of considering the temporal dependence of sequential driving data that can be used to estimate the AV's behavior from the human viewpoint. In the GRU model, we consider four input features to predict the future acceleration of the AV over the prediction horizon $N$, such as the features of *gap distance* $d_{\mathcal{M}_\mathcal{H}}$, *relative speed* $rs_{\mathcal{M}_\mathcal{H}}$ and *time headway* between the human driver and the AV $\tau_{\mathcal{M}_\mathcal{H}}$, and the *acceleration* of the AV $u_{\mathcal{M}_\mathcal{R}}$. The input states of the GRU model at the time $t$ can be described as:

$$x^t_G = \left\{d^t_{\mathcal{M}_\mathcal{H}}, rs^t_{\mathcal{M}_\mathcal{H}}, \tau^t_{\mathcal{M}_\mathcal{H}}, u^t_{\mathcal{M}_\mathcal{R}}\right\}, \quad X^t_G = \left[x^{t-1}_G, x^{t-2}_G, ..., x^{t-H}_G\right] \quad (6)$$

where $H$ is the historical time horizon. The output states of the GRU model at the time $t$ can be expressed as:

$$Y^t_G = \left[u^t_{\mathcal{M}_\mathcal{R}}, u^{t+1}_{\mathcal{M}_\mathcal{R}}, ..., u^{t+N-1}_{\mathcal{M}_\mathcal{R}}\right] \quad (7)$$

The choice of the features is made based on the feature importance analysis method in [12] and replicating the information on which the human driver is likely to base his/her driving decisions. At last, the length of the historical time horizon $H$ in the GRU prediction model is set as 10 s [11-12]. To train the GRU prediction model, we use the simulated trajectories with the trust-aware control design featuring the constant acceleration method (TAC-CA) and the simulation study details will be later presented in Section V.

**Constant motion prediction:** We use the constant acceleration model as a baseline plan estimation method to describe the future motion of the AV's plan in the human driver's mind. The constant acceleration model is one of the most widely used prediction models in short-horizon motion prediction applications [14-15].

### C. Trust Dynamics Modeling

In this study, we model the human driver's trust level as a plan assessment metric for determining the trust in human-AV interaction to reflect the human's cognitive trust level originating in the human's mind that is difficult to quantify. Specifically, we consider a plan evaluation factor that quantifies the explicability of the AV's plan from a human's perception to construct the trust level, as motivated from [16]. First, we formulate the explicability score of the AV's plan as the *dist* function in (4). This *dist* function is formulated as the action distance between the AV's original plan $\mathbf{u}_{\mathcal{M}_\mathcal{R}}$ and the expected AV's plan from the human drivers' perspective $\mathbf{u}_{\mathcal{M}^{\mathcal{H}}_\mathcal{R}}$ by using the Jaccard similarity coefficient, a commonly used method to measure the distances between plans in human-robot interactions [17-18]:

$$E(t) = dist\left(\mathbf{u}_{\mathcal{M}_\mathcal{R}}, \mathbf{u}_{\mathcal{M}^{\mathcal{H}}_\mathcal{R}}\right) = \frac{\left|\mathbf{u}_{\mathcal{M}_\mathcal{R}} \cap \mathbf{u}_{\mathcal{M}^{\mathcal{H}}_\mathcal{R}}\right|}{\left|\mathbf{u}_{\mathcal{M}_\mathcal{R}} \cup \mathbf{u}_{\mathcal{M}^{\mathcal{H}}_\mathcal{R}}\right|}$$
$$= \begin{cases} 0, & \text{if } \mathbf{u}_{\mathcal{M}_\mathcal{R}} = \mathbf{u}_{\mathcal{M}^{\mathcal{H}}_\mathcal{R}} \\ \dfrac{\sum_{k=0}^{N-1}\left(u^k_{\mathcal{M}_\mathcal{R}} + \varepsilon\right)\left(u^k_{\mathcal{M}^{\mathcal{H}}_\mathcal{R}} + \varepsilon\right)}{\sum_{k=0}^{N-1}\left(\left(u^k_{\mathcal{M}_\mathcal{R}} + \varepsilon\right)^2 + \left(u^k_{\mathcal{M}^{\mathcal{H}}_\mathcal{R}} + \varepsilon\right)^2 - \left(u^k_{\mathcal{M}_\mathcal{R}} + \varepsilon\right)\left(u^k_{\mathcal{M}^{\mathcal{H}}_\mathcal{R}} + \varepsilon\right)\right)}, & \text{else} \end{cases}$$
(8)

where $E(t)$ defines the explicability score of the AV's plan at the time step $t$ and $\varepsilon$ is the positive acceleration offset value to avoid the negative action distance calculation in the *dist* function. Second, we define the performance level of AV by incorporating the explicability score defined in (8):

$$P(t) = 1 - \tanh\left(E(t)\right) \quad (9)$$

where $P(t)$ denotes the performance level of the AV at the time step $t$ where $P(t) \in [0, 1]$. The intuitive explanation of the two extreme performance levels of the AV is that the AV achieves its highest performance level (level 1) when the AV's plan is perfectly explicable for the human driver. On the other hand, the performance level of the AV approaches its lowest performance level (level 0) when the explicability of the AV's plan reduces for the human driver.

Third, we formulate the trust level by incorporating the performance level of the AV as defined in (9):

$$T(t+1) = T(t) + \delta u_T(t) \quad (10)$$

where $T(t)$ and $u_T(t)$ are the trust level and trust input at the time step $t$, respectively, and $\delta$ is a positive constant to describe the growth rate of the trust. The growth rate of the trust level $\delta$ is set to 0.1, considering the dynamics of the human driver's trust level is slow in practice, as discussed in [6]. The trust input is defined as:

$$u_T(t) = \begin{cases} u_T(t) = P(t), & \text{if } P(t) >= P_{\text{thre}} \\ u_T(t) = -(1-P(t)), & \text{else} \end{cases} \quad (11)$$

where $P_{\text{thre}}$ is the threshold of the performance level in which the current trust level starts increasing. We assume that $P_{\text{thre}} = 0.8$ by referring to existing studies in human-robot trust modeling [6], [19-20].

IV. TRUST-AWARE CONTROL DESIGN

We formulated the trust-aware behavior planning of the AV in the previous section. Next, we will introduce the trust-aware control design for the AV to build its longitudinal plan during the operation of the vehicle. By incorporating the preview information from the preceding traffic (PT), the proposed approach minimizes the cost function (4) across the prediction horizon. Vehicle connectivity and sophisticated sensing technologies can be used to gather such preview information [1].

A. Trust-Aware Control Design Formulation

In this study, a nonlinear MPC (NMPC) algorithm is proposed for solving the AV's trust-aware control approach. At each time step $t$, the NMPC design aims to generate the optimal longitudinal acceleration vector of the AV with respect to the cost function and the system constraints for every prediction step $k$. The longitudinal speed $v_{\mathcal{M}_{\mathcal{R}}}$ is constrained by a predetermined minimum and maximum speed based on the traffic conditions. The AV's car-following gap distance is $d_{\mathcal{M}_{\mathcal{R}}}$ constrained by the minimum and maximum values considering safety clearance and reasonable distance for vehicle connectivity, respectively. The longitudinal acceleration $u_{\mathcal{M}_{\mathcal{R}}}$ constraints are applied to ensure the vehicle's drivability. Finally, the constraint on the trust level $T$ is applied to maintain the trust levels higher than the predefined minimum trust level. The NMPC design parameters are listed in Table I. The AV's optimal control problem can be stated as follows:

$$\mathbf{u}^*_{\mathcal{M}_{\mathcal{R}}} = \arg\min_{\mathbf{u}_{\mathcal{M}_{\mathcal{R}}}} \left( C_l\left(x^t_{\mathcal{M}_{\mathcal{R}}}, \mathbf{u}_{\mathcal{M}_{\mathcal{R}}}\right) + \alpha \cdot dist\left(\mathbf{u}_{\mathcal{M}_{\mathcal{R}}}, \mathbf{u}_{\mathcal{M}^{\mathcal{H}}_{\mathcal{R}}}\right) \right)$$

$$\text{s.t. } d_{\mathcal{M}_{\mathcal{R}_{\min}}} \leq d^k_{\mathcal{M}_{\mathcal{R}}} \leq d_{\mathcal{M}_{\mathcal{R}_{\max}}}, \quad v_{\mathcal{M}_{\mathcal{R}_{\min}}} \leq v^k_{\mathcal{M}_{\mathcal{R}}} \leq v_{\mathcal{M}_{\mathcal{R}_{\max}}} \quad (12)$$

$$u_{\mathcal{M}_{\mathcal{R}_{\min}}} \leq u^k_{\mathcal{M}_{\mathcal{R}}} \leq u_{\mathcal{M}_{\mathcal{R}_{\max}}}, \quad T_{\min} \leq T^k$$

B. Human Driver Model

In this study, we use our previously developed inverse reinforcement learning (IRL) based driver behavior model [21] to estimate the longitudinal motion of the human driver. IRL-based driver behavior models have been widely used to understand the driving preferences of human drivers in real-world driving scenarios [21-23]. The objective of the IRL approach is to learn an inherent cost function that captures the observed driving preferences of the human driver in driving demonstrations. Here, we learn that the cost function of the human driver via the developed IRL approach with a given set of driver demonstrations which we collected from the driver-in-the-loop simulator [22], and the NMPC algorithm is used to generate driver-specific actions. The cost function and the constraints can be formulated as:

$$C_{\mathcal{M}_{\mathcal{H}}} = \mathbf{W}^T_{\mathcal{M}_{\mathcal{H}}} \sum_{k=0}^{N-1} \left( \mathbf{f}_{\mathcal{M}_{\mathcal{H}}} \left( x^{t,k}_{\mathcal{M}_{\mathcal{H}}}, \mathbf{u}^k_{\mathcal{M}_{\mathcal{H}}}, \mathbf{u}^k_{\mathcal{M}^{\mathcal{H}}_{\mathcal{R}}} \right) \right)$$

$$\mathbf{f}_{\mathcal{M}_{\mathcal{H}}} = \left(f_a, f_{ds}, f_{rs}, f_{rd}\right)^T \quad (13)$$

$$\text{s.t. } d_s \leq d^k_{\mathcal{M}_{\mathcal{H}}}, \quad v_{\mathcal{M}_{\mathcal{H}_{\min}}} \leq v^k_{\mathcal{M}_{\mathcal{H}}} \leq v_{\mathcal{M}_{\mathcal{H}_{\max}}}$$

where $\mathbf{f}_{\mathcal{M}_{\mathcal{H}}}$ is the feature vector which consists of *acceleration*, *desired speed*, *relative speed*, and *relative distance* features, and $\mathbf{W}_{\mathcal{M}_{\mathcal{H}}}$ is the learned feature weight vector. The longitudinal speed $v_{\mathcal{M}_{\mathcal{H}}}$ is constrained by predefined minimum and maximum speed values. The reader is referred to [21] for more information regarding the driver behavior model.

Table I: NMPC design parameters.

| Parameter | Value | Parameter | Value |
|---|---|---|---|
| $N$ | 3 s | $v_{\mathcal{M}_{\mathcal{R}_{\max}}}, v_{\mathcal{M}_{\mathcal{H}_{\max}}}$ | $\max(v_{PT})$ |
| $u_{\mathcal{M}_{\mathcal{R}_{\min}}}$ | -3 m/s² | $v_{\mathcal{M}_{\mathcal{R}_{\min}}}, v_{\mathcal{M}_{\mathcal{H}_{\min}}}$ | 0 m/s |
| $u_{\mathcal{M}_{\mathcal{R}_{\max}}}$ | 3 m/s² | $T_{\min}$ | 0 |
| $d_{\mathcal{M}_{\mathcal{R}_{\min}}}$ | 5 m | $d_{\mathcal{M}_{\mathcal{R}_{\max}}}$ | 60 m |
| $d_s$ | 5 m | $\tau_{\mathcal{M}_{\mathcal{R}}}, \varepsilon$ | 1.2 s, 6 m/s² |

V. RESULTS AND DISCUSSIONS

By utilizing the previously introduced models, we aim to explore the performance of the proposed trust-aware control design in human-AV car-following interactions. We specify five different relative weight values of the AV's trust-aware control design for the comparison study where the relative weight values are described as $\alpha \in [0, 0.25, 0.50, 0.75, 1.00]$. Besides, we analyze each comparison case with the GRU and constant acceleration motion prediction models to effectively investigate the prediction performance of these plan estimation methods and their effectiveness with the proposed system. For the sake of conciseness, trust-aware control design with the GRU and constant acceleration prediction models are denoted as "TAC-GRU" and "TAC-CA" in the graphical and numerical results, respectively.

We use two standard driving cycles that are US06 and New York City Cycle (NYCC) [24] to validate the performance of the proposed method with the commonly known standardized highway and urban driving cycles. US06 and NYCC driving cycles are adopted by the US Environmental Protection Agency (US EPA) and represent highly dynamic vehicle trajectories of the highway and urban driving scenarios, respectively. We further verify the performance of the proposed design with two experimental vehicle trajectories that we collected from daily commuting traffic that consists of highway and urban driving scenarios. More details about the experimental data collection can be found in [21]. The speed profiles in these vehicle trajectories are used to define the PT's speed profile to simulate the realistic preceding traffic in front of the human-AV interaction. For illustration purposes, these speed profiles are plotted in Fig. 2.

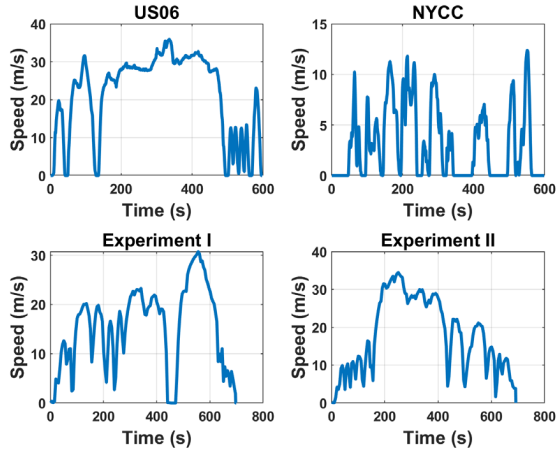

Fig. 2  Speed profiles of the preceding traffic.

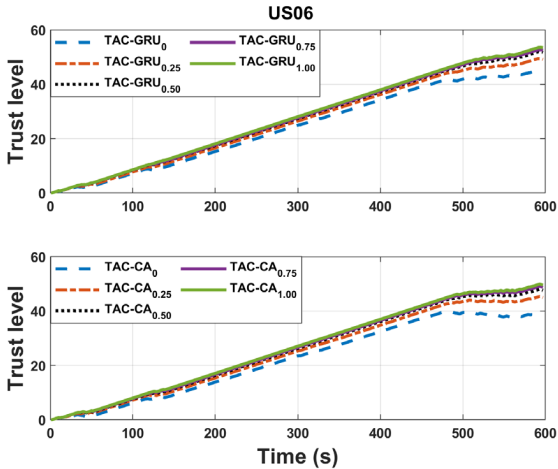

Fig. 3  Trust level of the human driver during the trip with TAC-GRU and TAC-CA approaches in the US06 testing.

We use the trust level, average explicability score, and root mean square error (RMSE) of the acceleration prediction models for each comparison case to assess the performance of the proposed trust-aware control designs.

We first evaluate the performance of the proposed TAC-GRU and TAC-CA approaches in the US06 testing. Fig. 3 shows the trust level of the human driver during the trip with the TAC-GRU and TAC-CA approaches. We find that the trust level increases faster when the relative weight $\alpha$ increases in both approaches. The fundamental reason for this is that the trust-awareness of the AV's plan in both approaches increases in parallel with the increase in the AV's effort to improve the explicability of its plan in human-AV interactions. According to Table II and Table III, we observe that as the relative weight $\alpha$ increases, that is, as the trust-awareness of the AV increases, the trust level of the human driver significantly increases and achieves a 20-30% maximum improvement in the proposed approaches. Besides, when the relative weight $\alpha$ increases, the average explicability score and the RMSE value of the AV's plan reduce and achieve 56-60% and 29-32% maximum improvements, respectively. These findings demonstrate that the trust-aware AV generates more explicable plans in which the human driver can better predict the AV's behaviors, and the human driver ends the trip with a significantly increased trust level towards the AV when the AV increases its effort to maximize its plan explicability during the trip.

Table II: Statistical comparison of the TAC-GRU approach with different relative weight values in US06 testing.

| Relative weight ($\alpha$) | Final Trust Level | Average Explicability Score | RMSE (m/s$^2$) |
|---|---|---|---|
| 0 (trust-unware) | 44.360 | 0.038 | 0.518 |
| 0.25 | 49.160 | 0.024 | 0.435 |
| 0.50 | 51.707 | 0.019 | 0.393 |
| 0.75 | 52.500 | 0.016 | 0.367 |
| 1.00 | 53.402 | 0.015 | 0.353 |
| **Maximum improvement (%)** | **20.38%** | **59.27%** | **31.87%** |

Table III: Statistical comparison of the TAC-CA approach with different relative weight values in US06 testing.

| Relative weight ($\alpha$) | Final Trust Level | Average Explicability Score | RMSE (m/s$^2$) |
|---|---|---|---|
| 0 (trust-unware) | 38.141 | 0.065 | 0.724 |
| 0.25 | 44.991 | 0.041 | 0.616 |
| 0.50 | 47.611 | 0.034 | 0.570 |
| 0.75 | 48.762 | 0.031 | 0.537 |
| 1.00 | 49.539 | 0.028 | 0.513 |
| **Maximum improvement (%)** | **29.88%** | **56.37%** | **29.22%** |

Table IV: Maximum improvement of the trust-aware behaviors with the TAC-GRU and TAC-CA approaches in the real-world vehicle trajectories testing.

| **TAC-GRU** | Final Trust Level | Average Explicability Score | RMSE (m/s$^2$) |
|---|---|---|---|
| NYCC | 20.12% | 66.55% | 39.33% |
| Experiment 1 | 9.72% | 72.78% | 45.25% |
| Experiment 2 | 9.89% | 73.81% | 45.16% |
| **TAC-CA** | Final Trust Level | Average Explicability Score | RMSE (m/s$^2$) |
| NYCC | 24.01% | 48.87% | 48.87% |
| Experiment 1 | 7.59% | 36.36% | 57.48% |
| Experiment 2 | 8.78% | 36.68% | 60.52% |

Furthermore, the numerical results in Table II and Table III indicate that the TAC-GRU approach provides a better final trust level and better plan explicability for the human driver compared to the TAC-CA approach for each comparison case. The primary reason for this is that the GRU acceleration prediction model has better prediction accuracy compared to the baseline constant acceleration prediction model to estimate the AV's plan from the human driver's perspective.

At last, we analyze the performance of the proposed designs with the previously mentioned other driving cycles to verify the generalizability of the proposed design. The numerical results in Table IV show that AV's trust-aware behaviors provide a maximum 7-24%, 36-74%, and 39-60% improvement in final trust level, average explicability score, and RMSE value in the proposed approaches, respectively. These results collectively demonstrate that the proposed trust-aware control designs have the potential to improve the trust level of the human driver by boosting the explicability of the AV's plan in different realistic human-AV car-following interactions.

## VI. CONCLUSION

In this study, we proposed a trust-aware control design for the automated vehicle to generate explicable and trust-aware plans for the following human driver in the car-following scenarios. The trust level of the following human driver is modeled as the plan evaluation index to evaluate the trustworthiness at the level of vehicle-to-vehicle interaction. The explicability of the AV's plan is then incorporated into the decision-making process of the AV to achieve trust-aware control. The statistical results indicate that the automated vehicle's trust-aware behaviors can considerably boost the explicability of the AV's plan and the following human driver's trust level in the car-following interactions.

The proposed approach is a first step in building a trust-aware behavior design for automated vehicles in the car following interactions with human drivers. We have assumed that the trust level can be modeled as the explicability of the AV's plan from human perception. We will expand on this work by designing alternative trust models in human-AV interactions and constructing real-time driver-in-the-loop simulations to analyze the performance of the proposed framework.